\documentclass[12pt]{article}
\usepackage{amsmath}
\usepackage{hyperref}
\usepackage{graphicx}
\usepackage[a4paper, margin=1in]{geometry}
\usepackage{soul}
\usepackage{xcolor}
\sethlcolor{yellow}
\usepackage{tikz}
\usepackage{amsmath}
\usepackage{authblk} 
\usepackage{lmodern}
\usepackage{fancyhdr}
\usepackage{titling}
\usepackage{fontawesome5} 
\usepackage{setspace}      

\usetikzlibrary{shapes.geometric, arrows.meta, positioning}
\usepackage{tikz}
\usetikzlibrary{shapes.geometric, arrows.meta, positioning, calc}

\tikzstyle{block} = [rectangle, draw, fill=gray!10, text width=4.5cm, text centered, rounded corners, minimum height=2.2cm]
\tikzstyle{arrow} = [thick,->,>=stealth]

\title{\textbf{Holographic Projection and Cyber Attack Surface: A Physical Analogy for Digital Security}}
\author{
    Ricardo Q. A. Fernandes$^{1}$, Anderson Santos$^{2,3}$,\\Daniel M. de Carvalho$^{1}$, André Luiz B. Molina$^{1}$\\\small $^{1}$Secretaria de Segurança da Informação e Cibernética (SSIC/GSI)\\
    \small $^{2}$Instituto Militar de Engenharia (IME)\\
    \small $^{3}$Venturus Centro de Inovação Tecnológica
    \\\scriptsize\texttt{\{ricardo.fernandes,daniel.maier,molina\}@presidencia.gov.br},\texttt{anderson@ime.eb.br}
}
\date{V JORNADAS STIC - Panamá 2025}

\begin{document}
\maketitle

\section*{Abstract}
This article presents an in-depth exploration of the analogy between the Holographic Principle in theoretical physics and cyber attack surfaces in digital security. Building on concepts such as black hole entropy and AdS/CFT duality, it highlights how complex infrastructures project their vulnerabilities onto their external interfaces. The paper draws a parallel between a black hole’s event horizon, which encodes all internal information, and the attack surface, which reflects the internal architecture’s security posture. Additionally, the article outlines how this conceptual framework can guide cybersecurity practices, emphasizing strategies such as attack surface reduction, continuous scanning with tools like OWASP ZAP and Greenbone OpenVAS, and the implementation of Zero Trust Architecture. This analogy not only provides a unique perspective on digital security but also underscores the critical importance of boundary-level defenses in protecting vast internal infrastructures.

\section{Introduction}
The Holographic Principle, first introduced by Gerard ’t Hooft \cite{tHooft1993} and later expanded by Leonard Susskind \cite{Susskind1995}, is a fundamental concept in theoretical physics suggesting that all information contained within a three-dimensional volume can be fully described by data encoded on its two-dimensional boundary. This principle finds significant application in the context of black hole physics, where it explains the proportionality between a black hole's entropy and the area of its event horizon, rather than its volume \cite{Bekenstein1973, Hawking1975}. This revolutionary idea has also influenced quantum gravity theories, particularly within the framework of the AdS/CFT correspondence \cite{Maldacena1998}. In this paper, we propose that the digital attack surface behaves analogously to a black hole’s event horizon, encoding the system’s internal vulnerabilities in a condensed, two-dimensional interface.
In the realm of cybersecurity, the concept of an \underline{attack surface} aligns remarkably with the Holographic Principle. The attack surface represents the total sum of all entry points through which an attacker can exploit vulnerabilities within a system. Like the event horizon of a black hole, the attack surface serves as the boundary where internal complexities become exposed to external threats. This surface encompasses not only technical endpoints, such as APIs, web services, and email servers, but also social and organizational elements, including user behavior, third-party access, and policy enforcement.
The attack surface is not static; it evolves with system updates, new deployments, and changes in user behavior. Minimizing it requires a multi-faceted approach, combining technical defenses with a strong security culture. This includes continuous monitoring, timely patching, least privilege access controls, and comprehensive user awareness training. Moreover, the rise of IoT and XIoT (Extended Internet of Things) devices further expands the attack surface, introducing new challenges in managing vulnerabilities across interconnected systems.
By drawing parallels between the Holographic Principle and cybersecurity, this paper demonstrates how understanding and securing the external attack surface can protect the underlying infrastructure. Just as the event horizon encodes the internal state of a black hole, the attack surface reflects the security posture of an entire digital ecosystem. This analogy provides a powerful framework for developing more effective cybersecurity strategies, emphasizing boundary-level defenses while addressing the human and organizational factors that shape the attack surface.

\section{Holographic Principle and Entropy}

The Holographic Principle finds its origin in black hole thermodynamics, where entropy—a measure of information—is proportional not to volume but to the area of the event horizon. This groundbreaking discovery suggests that all information about the internal state of a black hole is encoded on its two-dimensional boundary. The relationship is expressed as:

\begin{equation}
\label{eq:entropy}
    S = \frac{k c^3}{\hbar G} \frac{A}{4}
\end{equation}

Where:
\begin{itemize}
    \item $S$: Entropy of the black hole (measure of hidden information)
    \item $A$: Area of the event horizon
    \item $k$: Boltzmann constant (relates entropy to information)
    \item $\hbar$: Reduced Planck constant (quantum scale factor)
    \item $G$: Gravitational constant
    \item $c$: Speed of light
\end{itemize}

This formula reveals that the entropy—and thus the maximal information content—of a black hole is not determined by the volume it occupies, but solely by the area of its horizon. This challenges conventional notions from classical physics and suggests a deeper informational structure underlying spacetime geometry.

The Holographic Principle, formalized by Gerard 't Hooft~\cite{tHooft1993} and later expanded by Susskind~\cite{Susskind1995}, generalizes this idea. It postulates that any region of space can be fully described by degrees of freedom residing on its boundary surface, analogous to how a hologram encodes three-dimensional information on a two-dimensional film.

In modern theoretical physics, this principle finds rigorous support in the Anti-de Sitter/Conformal Field Theory (AdS/CFT) correspondence~\cite{Maldacena1998}, one of the most profound results in string theory. The AdS/CFT correspondence states that a gravitational theory defined in a $(d+1)$-dimensional Anti-de Sitter (AdS) space is mathematically equivalent to a conformal field theory (CFT) without gravity defined on the $d$-dimensional boundary of that space. In other words, the bulk gravitational dynamics—including black holes—can be fully described by a quantum field theory living on the boundary. This duality is not merely metaphorical, but a precise mathematical equivalence, offering a powerful framework for studying quantum gravity and strongly coupled quantum systems.

The AdS/CFT correspondence provides a concrete realization of the Holographic Principle by showing that bulk physics—including information content, thermodynamics, and causality—can be projected and recovered entirely from the lower-dimensional boundary theory. This insight forms the backbone of many modern explorations in quantum gravity, entanglement entropy, and black hole information paradox resolution.

An important consequence of this principle emerges when a black hole absorbs additional matter or radiation. As new information enters the black hole, the area of its event horizon increases. This expansion ensures that the total entropy also increases, maintaining the thermodynamic consistency of the system. In other words, the holographic projection adapts in real-time to incorporate the additional information, extending the event horizon accordingly. Each bit of incoming data contributes to the expansion of the two-dimensional boundary, while the internal complexity remains hidden beneath it.

This concept—that the informational content of a spatial volume is encoded on its boundary—has profound implications not only for quantum gravity but also for disciplines seeking to interpret systems through their observable interfaces, as we explore in the cybersecurity analogy proposed in this article.

\section{What is an Attack Surface?}
In cybersecurity, the \textbf{attack surface} encompasses all possible points where an unauthorized user could access a system and extract data. This includes hardware, software, network components, and human elements that could be exploited \cite{ibm_attack_surface}. Understanding and defining this surface is challenging due to the complex and interconnected nature of \textbf{security boundaries} across different perspectives.

Security boundaries are barriers established to protect resources and information from unauthorized access. Depending on the perspective, these boundaries vary:

\begin{itemize}
    \item \textbf{Physical}: Controls access to facilities and physical devices.
    \item \textbf{Digital}: Involve firewalls, intrusion detection systems, and measures protecting networks and data.
    \item \textbf{Computational}: Encompasses security mechanisms within operating systems and applications.
    \item \textbf{Virtual}: Pertain to virtual and cloud environments, where boundaries are defined by software configurations.
    \item \textbf{Social}: Relates to human behavior, organizational policies, and security culture.
\end{itemize}

These security boundaries are not necessarily nested or hierarchical. Often, they overlap, creating hybrid zones that combine internal and external elements from different boundaries. For instance, an IoT device is a physical asset with digital interfaces, operating in a virtual environment, and managed by computational and social policies. This interconnection complicates the clear definition of the attack surface, as a vulnerability in one boundary can affect others.

Due to the overlap and interdependence of security boundaries, mapping and defining the attack surface becomes complex. The expansion of digital infrastructures, adoption of IoT devices, and increasing interconnectivity amplify potential entry points. Additionally, the human factor introduces unpredictable variables, such as errors or negligent behaviors, further expanding the attack surface.

Considering these aspects, the attack surface can be defined as:

\begin{quote}
    \textit{"The total set of entry points, encompassing physical, digital, computational, virtual, and social dimensions, through which unauthorized agents may attempt to access or compromise the integrity, confidentiality, or availability of an organization's resources and information."}
\end{quote}

In cybersecurity, \textbf{entropy} can be understood as a measure of disorder or unpredictability within a system. An extensive and complex attack surface indicates high entropy, suggesting a greater number of vulnerabilities and potential attack paths. For example:

\begin{itemize}
    \item \textbf{Physical}: Environments with multiple physical access points and dispersed devices increase entropy due to monitoring and control challenges.
    \item \textbf{Digital}: Systems with numerous applications, exposed services, and APIs expand the digital attack surface, elevating entropy.
    \item \textbf{Computational}: Utilizing diverse operating systems and software, especially without regular updates, contributes to disorder and vulnerabilities.
    \item \textbf{Virtual}: Poorly configured virtual environments or those with excessive permissions introduce additional complexity, increasing entropy.
    \item \textbf{Social}: Lack of security training and inconsistent policies can lead to unpredictable behaviors, raising the risk of social engineering exploitation.
\end{itemize}

Reducing entropy in each of these perspectives through measures such as strict access control, regular updates, network segmentation, and continuous training is essential to minimize the attack surface and strengthen an organization's security posture.

In the realm of cybersecurity, the attack surface represents the set of all possible entry points through which an attacker may exploit vulnerabilities within a system. Similar to how the Holographic Principle suggests that all three-dimensional information in a physical system is encoded on its two-dimensional boundary, the security posture of an entire digital infrastructure can be understood by analyzing its external exposure.

A system's complexity—consisting of databases, applications, network protocols, authentication mechanisms, and cloud services—forms an intricate internal structure. However, attackers do not need direct access to this full complexity; they only need to interact with its externally exposed interfaces, much like how information in a black hole is accessible only at the event horizon. The attack surface, therefore, acts as a holographic projection of the internal system, revealing the pathways that an adversary can leverage to compromise security.

From a security standpoint, the broader and more complex the attack surface, the higher the likelihood of successful exploitation. Every open API, web application, exposed endpoint, and misconfigured cloud service increases the entropy of the system, introducing more potential avenues for attack. This is analogous to how an increase in the area of a black hole’s event horizon corresponds to an increase in its entropy, representing greater informational complexity that can be manipulated or exploited.

Minimizing the attack surface is one of the most effective defensive strategies in cybersecurity. This involves:
\begin{itemize}
    \item Reducing externally accessible services by disabling unnecessary ports and interfaces.
    \item Implementing rigorous authentication and authorization mechanisms to ensure that access is restricted to trusted entities.
    \item Employing continuous security monitoring and scanning with tools like OWASP ZAP and Greenbone OpenVAS to identify and patch vulnerabilities before they can be exploited.
    \item Applying Zero Trust Architecture (ZTA), which assumes that no entity should be inherently trusted, ensuring strict verification at every level.
    \item Segmenting networks and enforcing least privilege principles to limit lateral movement within an infrastructure in case of a breach.
\end{itemize}

By treating the attack surface as a holographic boundary that encodes all exploitable interactions with a system, security professionals can prioritize defenses at the perimeter while maintaining internal resilience. Just as in black hole physics, where information loss is a fundamental concern, data breaches in cybersecurity stem from how well—or poorly—an organization protects its informational boundary. A well-secured attack surface ensures that even if adversaries observe the system’s perimeter, they gain minimal actionable insight into its internal structure, preventing deeper exploitation.

\section{The Analogy Between the Holographic Principle and Cybersecurity}
The physical concept of the Holographic Principle states that all three-dimensional information of a system is encoded on its two-dimensional surface. This serves as a powerful metaphor for understanding the relationship between the cyber attack surface and the volume of underlying infrastructures and services. Just as a black hole's entropy, which measures its information content, is proportional to its event horizon's area rather than its volume, the risk and exposure of a system to cyber threats depend more on its external, exposed attack surface rather than its total complexity. 

The attack surface represents the sum of all accessible interfaces, endpoints, APIs, and network entry points through which adversaries can attempt to exploit vulnerabilities. Meanwhile, the underlying infrastructure, composed of databases, internal networks, and protected resources, forms the hidden bulk of the system, much like the unseen interior of a black hole. This analogy underscores the importance of focusing cybersecurity efforts on minimizing and securing the external attack surface to protect the vast internal architecture it represents.

For instance, in an e-commerce infrastructure, exposed public APIs can be viewed as part of the system’s "holographic surface". An attacker may exploit one of these APIs through an SQL injection — without requiring direct access to the underlying database — demonstrating how a vulnerability at the boundary can lead to a compromise of the entire system. This scenario exemplifies how the attack surface, much like the event horizon of a black hole, acts as the interface through which internal complexity becomes exposed and exploitable.

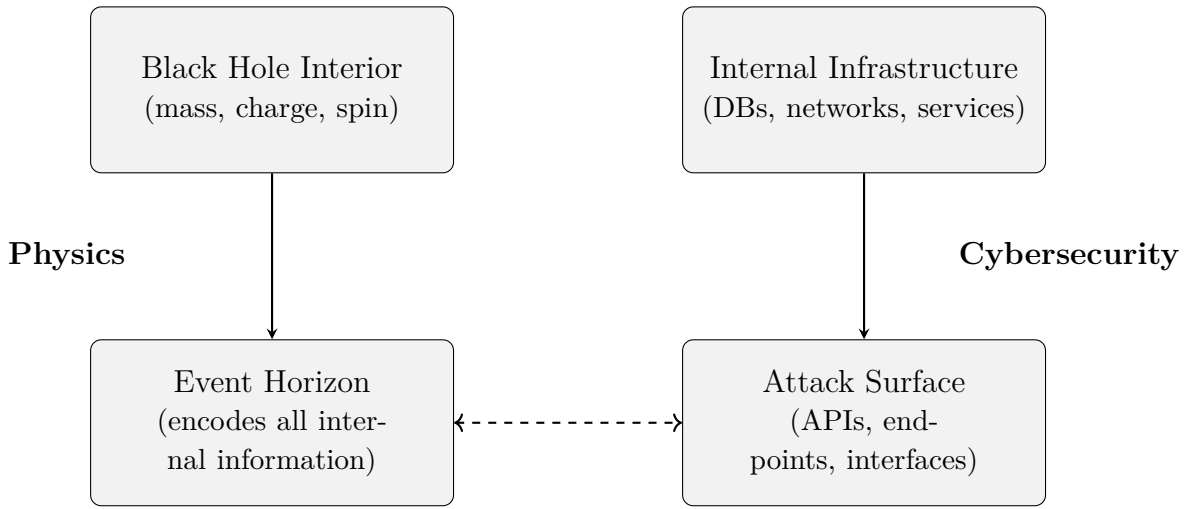
\begin{figure}[h!]
\centering
\begin{tikzpicture}[node distance=2.2cm and 3cm]

\node[block] (bh_interior) {Black Hole Interior\\\small (mass, charge, spin)};
\node[block, below=of bh_interior] (bh_horizon) {Event Horizon\\\small (encodes all internal information)};
\node at ($(bh_interior)!0.5!(bh_horizon) + (-2.7,0)$) {\textbf{Physics}};
\draw[arrow] (bh_interior) -- (bh_horizon);

\node[block, right=of bh_interior] (sys_internal) {Internal Infrastructure\\\small (DBs, networks, services)};
\node[block, below=of sys_internal] (attack_surface) {Attack Surface\\\small (APIs, endpoints, interfaces)};
\node at ($(sys_internal)!0.5!(attack_surface) + (2.7,0)$) {\textbf{Cybersecurity}};
\draw[arrow] (sys_internal) -- (attack_surface);

\draw[<->, thick, dashed] (bh_horizon.east) -- (attack_surface.west) node[midway, above, sloped] {};

\end{tikzpicture}
\caption{Analogy between the Holographic Principle and Cybersecurity}
\label{fig:holography-cyber}
\end{figure}

Figure~\ref{fig:holography-cyber} illustrates the conceptual analogy between the Holographic Principle in physics and the notion of the attack surface in cybersecurity. On the left, the structure of a black hole is depicted, where the event horizon encodes all information about the black hole's internal states—such as mass, charge, and spin—despite being a two-dimensional boundary. This reflects the core idea of the Holographic Principle: that a system’s complexity can be fully represented on its surface.

On the right side, a parallel is drawn with a digital system, where the internal infrastructure—composed of databases, networks, and services—is shielded by an external boundary known as the attack surface. This surface includes accessible elements such as APIs, interfaces, and endpoints, which adversaries may target. Just as the event horizon reveals the informational content of the black hole, the attack surface reveals the system’s security posture.

The dashed line connecting both sides emphasizes the analogy: in both domains, the boundary layer acts as a compressed representation of internal complexity. This highlights the strategic importance of securing boundary interfaces to protect the entire structure they enclose.

On the other hand, the dynamic behavior of Black Hole entropy and, therefore, of the holographic projection over the Event Horizon parallels modern cybersecurity concerns. As digital systems grow in complexity—adding services, endpoints, or users—their attack surfaces expand. Just as the black hole's event horizon enlarges to encode more internal information, a digital infrastructure’s external exposure increases to accommodate new functionality, potentially introducing new vulnerabilities.

Understanding this correspondence reinforces the critical importance of managing growth and complexity in secure systems. Proactive monitoring and minimization of the attack surface act like controlling the area of the event horizon, preventing entropy—and risk—from growing unchecked.

In simple terms, the equation \ref{eq:entropy} shows that the greater the "surface" of the system (i.e., the area of the event horizon), the greater its internal complexity. In the context of digital security, this translates to: the more exposed interfaces a system has, the higher its risk of being attacked. Additionally, it implies that a black hole's complexity (its internal states) is fully represented on its surface. Conceptually, it introduces the idea that boundaries hold crucial security-relevant information—a core analogy to cybersecurity.

In cybersecurity, the attack surface acts similarly to the event horizon, representing the interface where internal complexities (services, databases, and processes) become vulnerable. The size and complexity of an infrastructure (volume) are less relevant to attackers than its exposure (surface). Just as the entropy of a black hole increases with surface area, the risk to a digital system increases with the size and complexity of its attack surface. Reducing the attack surface, therefore, is analogous to limiting the event horizon through security controls, such as access restrictions, endpoint protection, and network segmentation.

Furthermore, the entropy concept reflects potential vulnerabilities: more interfaces, endpoints, and APIs result in higher entropy and more opportunities for exploitation. Tools such as continuous scanning (e.g., OWASP ZAP, Greenbone OpenVAS) act as measurement instruments to map this 'entropy'—highlighting exposed weaknesses that could be exploited by adversaries.

This analogy underscores a central principle of cybersecurity: reducing exposure at the surface level is as crucial as protecting the internal systems. By controlling access points and monitoring external interfaces—like managing the event horizon—security professionals can prevent internal exploits, even when adversaries have full visibility of the system's boundary.

\subsection{AdS/CFT Duality and Security Boundaries}
The Anti-de Sitter/Conformal Field Theory (AdS/CFT) correspondence is a fundamental result in theoretical physics, describing a duality between gravity in a volume (AdS space) and a quantum field theory on its boundary (CFT). This duality provides a powerful analogy for understanding security boundaries in cybersecurity. In the AdS/CFT framework:
\begin{itemize}
    \item AdS (Volume): Represents the internal structure of the system, including databases, microservices, containers, and internal networks.
    \item CFT (Boundary): Represents the system's external interfaces, such as APIs, endpoints, firewalls, and user interfaces.
\end{itemize}

The duality implies that the behavior and security of the entire internal infrastructure (volume) can be inferred and controlled from the boundary, similar to how all information about AdS space is encoded on the CFT boundary. This correspondence maps directly onto key cybersecurity concepts:
\begin{itemize}
    \item Volume (AdS) $\rightarrow$ Internal Infrastructure: The complex and layered systems behind the attack surface, including CI/CD pipelines, databases, and internal services.
    \item Boundary (CFT) $\rightarrow$ Attack Surface: The digital perimeter that adversaries interact with, such as web interfaces, exposed ports, and APIs.
    \item Bulk-to-Boundary Mapping $\rightarrow$ Security Controls: The way internal vulnerabilities manifest as external weaknesses and how effective boundary defenses can prevent internal exploits.
\end{itemize}

Implications for Cybersecurity Strategy:
\begin{enumerate}
    \item Perimeter Defense as a Priority: Just as the CFT boundary encodes AdS dynamics, the attack surface reflects the system’s vulnerabilities. Strong firewall policies, API gateways, and web application firewalls (WAFs) are essential.
    \item Zero Trust Architecture: Aligning with the idea that all interactions occur through the boundary, Zero Trust ensures every access attempt is authenticated and authorized at the perimeter.
    \item Behavioral Analysis (Bulk Observables from Boundary): Continuous monitoring of boundary activity with Security Information and Event Management (SIEM) tools helps infer internal compromise attempts, similar to how boundary fields describe bulk properties in AdS/CFT.
    \item Incident Containment and Lateral Movement Prevention: Network segmentation limits an attacker’s ability to move from the boundary into the core, much like the AdS space is fully constrained by its boundary.
\end{enumerate}

The AdS/CFT duality offers a profound analogy for cybersecurity: the system’s attack surface (CFT boundary) encodes the security state of its entire internal infrastructure (AdS volume). By focusing on boundary defenses and continuously monitoring external interactions, cybersecurity professionals can prevent internal exploits and reduce risk. As in physics, understanding the boundary reveals everything about the system.

\section{Entropy and the Limits of Surface-Based Defense}

According to the analogy proposed in this article, the cyber attack surface of a digital system functions as a holographic boundary that reflects the system’s internal informational complexity. Just as in black hole physics the entropy—and hence the information content—is projected on the event horizon, in cybersecurity, the attack surface projects the internal vulnerabilities of an organization’s infrastructure.

This leads to a critical insight: it is theoretically impossible to reduce the size of the attack surface without also reducing the entropy of the system. In other words, purely defensive strategies focused on the boundary—such as perimeter firewalls, API gateways, or access controls—can contain or obscure vulnerabilities, but they do not eliminate the underlying complexity that gives rise to the attack surface in the first place. In thermodynamic terms, entropy reflects disorder, variability, and uncertainty. In digital infrastructures, this entropy manifests through:
\begin{itemize}
    \item Redundant or outdated systems and services;
    \item Poorly documented APIs and endpoints;
    \item Uncontrolled access policies and legacy permissions;
    \item Inconsistent update and patching routines;
    \item Decentralized or fragmented security governance.
\end{itemize}

Reducing the attack surface in a sustainable way requires addressing the sources of entropy within the system. This means simplifying architectures, standardizing configurations, consolidating services, and eliminating unnecessary complexity. The application of Zero Trust Architecture~\cite{NIST800-207,ZTA2020}, continuous asset inventory, and strict lifecycle governance are practical ways to align internal entropy reduction with external surface minimization.

This principle echoes a fundamental law of black hole thermodynamics: surface area and entropy are intrinsically linked. Applying this insight to cybersecurity suggests that effective defense cannot rely solely on external hardening — it must be accompanied by strategic efforts to reduce internal disorder. Only then can the digital “event horizon” contract, making the system not just more opaque to attackers, but fundamentally less vulnerable.

\section{Discussion: Limits and Future Directions}
The analogy between holographic principles and cybersecurity, though insightful, has inherent limitations. One key limitation is the difference in dimensionality: the Holographic Principle is rooted in the nature of physical space-time, while attack surfaces in cybersecurity are abstract, multi-layered, and influenced by human behavior, policies, and dynamic system architectures.

Additionally, while entropy in black holes is a well-defined physical quantity, measuring entropy in cybersecurity remains an evolving field. Concepts such as system complexity, vulnerability density, and exploit probability are approximations that do not fully capture the thermodynamic rigor of the holographic model.

Potential for Future Research:
\begin{itemize}
    \item Theoretical Advancements: Further exploration of AdS/CFT duality-inspired models to simulate attack surface behaviors, aiding in the prediction and visualization of security threats.
    \item Entropy Observations in Cyber Systems: Developing quantifiable entropy models for digital infrastructures, which could validate or refute parallels to physical entropy models.
    \item Machine Learning and Quantum Computing: Leveraging advanced AI and quantum algorithms to detect patterns analogous to quantum states, providing deeper insights into attack surface dynamics.
    \item Cross-disciplinary Collaboration: Fostering collaborations between physicists and cybersecurity experts to explore how principles from quantum information theory can enhance security models.
\end{itemize}

The holographic analogy provides a novel lens for analyzing cybersecurity challenges. However, fully realizing its potential requires further research into entropy models, boundary behaviors, and interdisciplinary approaches. Future advancements in both theoretical physics and cyber analytics could bridge these domains, offering predictive tools and enhancing digital defense strategies.

\section{Conclusion}

This article proposed a novel analogy between the Holographic Principle in theoretical physics and the cyber attack surface in digital infrastructures. By drawing from black hole thermodynamics, the AdS/CFT correspondence, and entropy modeling, we built a conceptual framework to interpret cybersecurity exposure as a boundary-level projection of internal system complexity.

Through this analogy, we demonstrated that:

\begin{itemize}
    \item \textbf{The attack surface functions as a holographic projection}, encoding all exploitable aspects of an infrastructure's internal state.
    \item \textbf{AdS/CFT duality offers a powerful lens} to understand how internal security posture can be inferred from boundary behavior.
    \item \textbf{Reducing the attack surface requires more than perimeter hardening}; it demands reducing internal entropy through simplification, standardization, and lifecycle control.
    \item \textbf{Cybersecurity entropy} is a useful conceptual tool to assess systemic risk across physical, digital, computational, virtual, and social domains.
\end{itemize}

We fulfilled the objectives outlined in the abstract by articulating a theoretically grounded and practically applicable analogy. Strategies such as Zero Trust Architecture, continuous scanning (e.g., with OWASP ZAP and Greenbone OpenVAS), and asset lifecycle management align with the insights derived from this holographic perspective.

While the analogy has limitations—especially regarding the mathematical rigor of entropy in cybersecurity—the insights it offers are valuable. It reframes the attack surface not just as a set of vulnerabilities to be defended, but as a reflection of the system’s internal informational disorder. Reducing this disorder, or entropy, is essential for sustainable security.

Future directions include the formalization of entropy metrics in cybersecurity, integration with AI-based threat modeling, and further interdisciplinary collaboration. Bridging physics and cybersecurity may yield not only novel metaphors but new methods for securing increasingly complex digital ecosystems.

\bibliographystyle{plain}


\end{document}